\documentclass[12pt,preprint]{aastex}
\def\edcomment#1{\iffalse\marginpar{\raggedright\sl#1\/}\else\relax\fi}
\reversemarginpar

\begin{document}
\title{Gravitational waves from gamma-ray bursts}
 \author{Maurice H.P.M. van Putten}
\affil{LIGO Project, NW17-161, 175 Albany St., Cambridge, MA 02139-4307}

\begin{abstract}
We present a mechanism for long bursts of gravitational radiation from 
Kerr black holes surrounded by a torus. These systems are believed to form 
in core-collapse of massive stars in association with gamma-ray bursts. The 
torus catalyzes black hole-spin energy mostly into gravitational radiation, 
with a minor output in winds, thermal and neutrino emissions. Torus winds
impact the remnant envelope of the progenitor star from within, which may
account for X-ray emission lines and leaves a supernova remnant. The 
frequency in gravitational radiation satisfies $f_{gw}=470$Hz
$(E_{SNR}/4\times 10^{51})^{1/2}(0.1/\beta)^{1/2}(7M_\odot/M_H)^{3/2}$,
where $E_{SNR}$ is the kinetic energy in the SNR, $M_H$ is the black hole
mass and $\beta\simeq0.1$ the initial ejection velocity, as detected in
GRB 011211. Ultimately, this leaves a black hole binary surrounded by a
SNR, which is conceivably illustrated by RX J050736-6847.8.
\end{abstract}

\section{Introduction}
There is increasing evidence for a GRB-supernova association with
 young massive stars. These events are probably associated with the
 prompt formation of a Kerr black hole in core-collapse
 (Woosley 1993). The dissipative
 timescale of spin-energy in the horizon of the black hole agrees with
 the durations of tens of seconds of long GRBs, and a small fraction of
 the baryon-free energy in angular momentum released along the axis of
 rotation is consistent with the inferred baryon-poor input to the
 observed (Frail et al., 2001) $E_\gamma\simeq3\times 10^{51}$erg in 
 GRB-energies (van Putten, 2001; van Putten \& Levinson, 2003).

 Kerr black holes produce luminous output in gravitational radiation,
 winds, thermal and neutrino emissions through a surrounding torus.
 These torus emissions are contemporaneous with forementioned minor output 
 in baryon-poor outflows along the axis of rotation of the black hole. 
 Gravitational radiation forms a major output of the system, representing 
 approximately 10\% of the rotational energy $E_{rot}$ of the black hole. 
 About $1-2\%E_{rot}$ is emitted in MeV-neutrinos. Both these emissions do 
 not affect the environment. The output in torus winds of about $1\%E_{rot}$
 impact the remnant stellar envelope from within with continuum radiation 
 and with kinetic energy. We recently suggested that this 
 continuum radiation from within may be the source of X-ray line-emissions,
 when the envelope has expanded and goes through a transition from 
 optically thick to optically thin. This proposal is similar but not
 identical to the reflection model, as discussed by (Lazzati et al., 2002) 
 where line-emission is excited by incidence of continuum radiation on 
 the surface of an optically thick slab. The efficiency of continuum-to-line
 emission is generally on the order of a few percent in 
 such reflection processes.

 Based on Lazzati et al. (2002), Ghisellini et al. (2002) discuss a 
 model-dependent analysis of required continuum-energies $E_c$ for 
 excitation of line-emissions in GRB 970508 (Piro et al., 1999), 
 GRB 970828 (Yoshida et al., 1999), GRB 991216 (Piro et al., 2000), 
 GRB GRB 000214 (Antonelli 2000) and GRB 011211 (Rees et al. 2002).
 This indicates energies $E_c\ge 4\times 10^{52}$erg in GRB 991216,
 pointing towards an energy reservoir in excess of that 
 required for the GRB-energies $E_\gamma$, assuming
 a canonical efficiency of kinetic energy-to-gamma rays of about $15\%$.
 This supports the notion that the GRB inner engine is processing
 other channels of emissions, in addition to and in excess of 
 baryon-poor input to GRB-afterglow emissions.

 The predicted long burst in gravitational radiation associated with
 GRB-supernovae (``hypernovae") may be detected by upcoming gravitational 
 wave-experiments. We mention broad band laser {interferometric instruments 
 LIGO (Abramowici1992), VIRGO (Bradaschia 1992), TAMA (Massaki 1991) and GEO 
 (e.g., Schutz \& Papa, 1999), and bar or sphere detectors presently under 
 construction. Calorimetry on this energy output provides a method for 
 testing the existence of Kerr black holes as objects in nature.

\begin{figure}
%\plotfiddle{f8.eps}{3in}{0}{60}{60}{-180}{-125}
\plotone{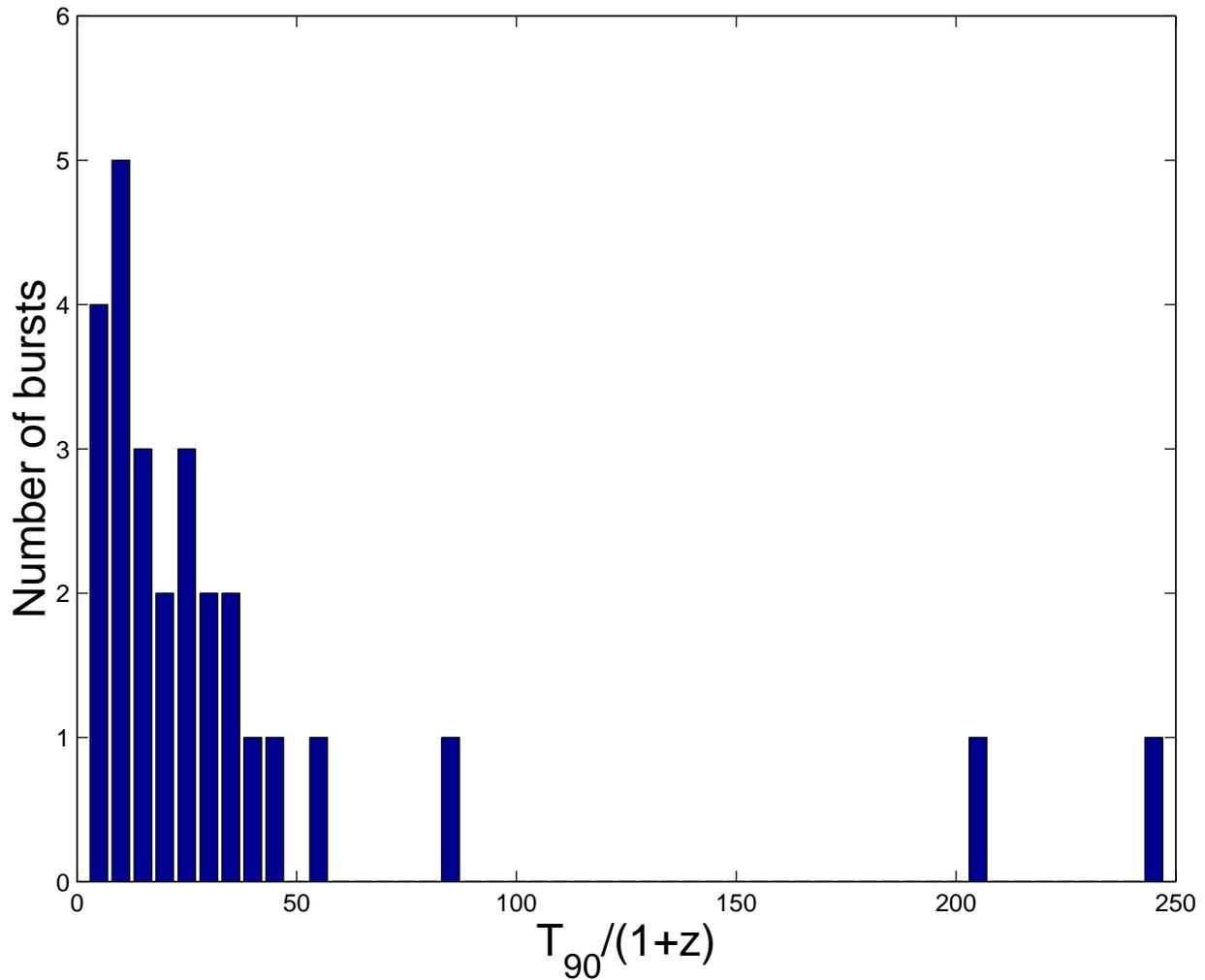}
\caption{Shown is the histogram of redshift corrected durations
of 27 long bursts with individually determined redshifts from 
their afterglow emissions. We identify these long durations with
the lifetime of rapid spin of a Kerr black hole in a state of
suspended accretion. These durations are effectively defined
by the rate of dissipation of black hole-spin energy in the
horizon -- an unobservable sink of energy -- subject to a new
magnetic stability criterion for the torus. (Reprinted from
M.H.P.M. van Putten, 2002, ApJ, 575, L71.)}
\end{figure}

 Calorimetry on hypernova remnants may further provide constraints 
 on the angular velocity of the torus and hence its frequency in 
 gravitational radiation. This serves to design and optimize specific 
 search strategies for the accompanying bursts of gravitational radiation. 
 Remnants of hypernovae are predicted to be black hole binaries surrounded by
 a supernova remnant. An example of this morphology is RX J050736-6847.8
 (Chu et al. 2001).

 In Section 2, we review radiation from Kerr black holes surrounded by
 a torus. In Section 3, we calculate a supernova connection from the 
 the impact of the torus winds on the remnant envelope of the progenitor
 star. We propose a remnant in Section 4, and conclude with suggestions
 for future observations in Section 5.

 \section{Black hole-beauty}
%Three-fold equivalence of tori to pulsars}

 A torus around a rotating black hole becomes radiant
 by {\em three-fold equivalence to pulsars}: 
 in poloidal topology, causality and ms rotation periods. 
 In accord with Mach's principle, the inner face of the torus 
 hereby receives input from the black hole -- 
 a nearby compact infinity with non-zero angular 
 velocity -- and the outer face radiates to asymptotic infinity. 
 In response, conservation of energy and angular momentum causes the 
 black hole to spin down, i.e.: {\em the black hole becomes luminous}. 
 See (van Putten \& Levinson, 2003) for a theory on tori surrounding
 rapidly spinning black holes.

 The torus develops a state of suspended accretion, while the black hole
 spins rapidly. The equations of suspended
 accretion permit solutions with a positive gravitational wave-luminosity,
 produced by multipole mass-moments in the torus.
 These emissions in gravitational radiation and winds are further accompanied
 by thermal and neutrino emissions, as dissipation in the torus develops
 MeV-temperatures. 

 Most of the spin-energy of the black hole is dissipated in the horizon. This
 dissipation is rate-limited, and sets a lower bound of tens of seconds on
 the lifetime of its spin and, hence, the emissions by the torus. Most of the
 black hole-luminosity is incident onto the torus, whereby the torus' emissions
 form the major energy output of the system. Gravitational radiation, in turn,
 defines most of the output from the torus, and represents about 10\% of the 
 spin-energy of the black hole.

 A small fraction of black hole-spin energy is released along the axis of
 rotation in the form of baryon-poor outflows along an open magnetic
 flux-tube. This magnetic flux-tube extends from the horizon to infinity.
 A horizon half-opening angle corresponding to the curvature in poloidal
 topology of the inner torus magnetosphere defines an energy output in
 baryon-poor outflows of about $0.1\%$ of the
 rotational energy of the black hole. This is in quantitative agreement
 with the observed value of $E_\gamma\simeq3\times 10^{51}$erg in GRB-energies,
 assuming a canonical efficiency of kinetic energy-to-gamma rays of about 15\%.

 We stress that our model for GRBs from rotating black holes 
 accounts quantitatively for both the 
 {\em long} durations relative to the Keplerian timescale (order parameter
 $\sim 10^4$), 
 as well as the {\em small} output in GRB-energies relative to the rotational 
 energy of the black hole (order parameter $\sim10^{-3}$).

 The model predicts a correlated output in gravitational radiation
 and torus winds.  In particular, their energies satisfy
 (van Putten (2003), corrected and simplified)
\begin{eqnarray}
\frac{E_{gw}}{E_{rot}}
    \simeq\frac{\alpha\eta}{\alpha(1+\delta)+f_w^2}\sim \eta
\label{EQN_EGW}
\end{eqnarray}
in the limit of strong viscosity (large $\alpha$)
and small slenderness (small $\delta$). Here $f_w$ denotes the
fraction of open magnetic flux supported by the torus which 
connects to infinity, $\delta=b/2a$ is a ratio of minor-to-major
radius of the torus and $\eta$ denotes the ratio of the angular 
velocity of the
torus to that of the black hole, which satisfies 
$\eta\sim 1/4\alpha$ for large $\alpha$. 
The energy emissions in winds satisfies 
\begin{eqnarray}
\frac{E_w}{E_{rot}}\simeq
    \frac{\eta f_w^2 (1-\delta)^2}{\alpha(1+\delta)+f_w^2}\sim \eta^2
\label{EQN_EW}
\end{eqnarray}
in the same limit considered in (1),
applied to the case of a symmetric flux-distribution with 
$f_w=\onehalf$. 
In dimensionful form, these energy emissions are
%\begin{eqnarray}
$E_{gw}\simeq4\times10^{53}\mbox{erg}\left({\eta}/{0.1}\right)
\left({M_H}/{7M_\odot}\right)$
%\end{eqnarray}
and
%\begin{eqnarray} 
$E_{w}\simeq4\times10^{52}\mbox{erg}\left({\eta}/{0.1}\right)^2
\left({M_H}/{7M_\odot}\right),$
%\end{eqnarray}
The frequency of quadrupole
gravitational radiation $f_{gw}$ and the energy in winds 
are related by
\begin{eqnarray}
f_{gw}\simeq 470 \mbox{Hz}
  \left(\frac{E_w}{4\times 10^{52}\mbox{erg}}\right)^{1/2}
  \left(\frac{7M_\odot}{M_H}\right)^{3/2}.
\end{eqnarray}
For the purpose of calorimetry on $E_w$, we set out to
identify observable signatures of the impact
of the torus winds on its surroundings.

\section{Black hole-supernovae}

Prompt core-collapse in young massive stars in binaries
may produce Kerr black holes surrounded by a disk or torus, formed from
matter stalled against an angular momentum barrier (Brown et al., 2000). 
Prompt collapse takes place on a 
free-fall time, or less through the agency of magnetic fields 
(see, e.g., van Putten \& Ostriker 2001).
As a result, the newly formed black hole-torus system is initially 
surrounded by a remnant stellar envelope. Spin-energy of the black hole 
released in the form of torus winds promptly impacts this stellar envelope 
from within.

 The kinetic energy and radial momentum of the torus winds 
 provides a powerful mechanism to eject 
 matter, producing a supernova associated with the underlying GRB.
 GRB 991216 and GRB 011211 both show initial ejection velocities $\beta\simeq 0.1$, 
 relative to the velocity of light, as observed in blue-shifted X-ray line-emissions.
 This blue-shift defines the efficiency $\beta/2$ of conversion of an energy $E_w$
 in essentially luminal torus winds to a kinetic energy $E_{SNR}$ of matter ejecta. 
 This leaves $E_{SNR}\simeq (1/2)\beta E_w\sim$ few
 $\times 10^{51}$ erg from $E_w=$ few $\times 10^{52}$ erg. These values
 are remarkably similar to the kinetic energies in non-GRB supernova 
 remnants. Because the efficiency $\beta$ of $E_w$ to $E_{SNR}$ 
 is somewhat larger than the efficiency of continuum-to-line
 emissions, it is a preferred method for calorimetry on torus 
 winds.

The energetic impact on the envelope provides a source of continuum
emission for excitation of X-ray lines, and deposits kinetic energy. 
The above shows that both of these processes are remarkably 
{\em inefficient.} Excitation 
of X-ray lines by continuum emissions has an estimated efficiency of 
less than one percent (Ghisellini 2002). Deposition of kinetic energy by 
approximately luminal torus winds has an efficiency of $\beta$, denoting 
the ejection velocity relative to the velocity of light.

Matter ejecta in both GRB 991216 (Piro et al., 2000) and
GRB 011211 (Reeves, 2002) show an expansion velocity of 
$\beta\simeq 0.1$.
The efficiency of kinetic energy deposition of the torus wind onto
this remnant matter is hereby $\beta/2=5\%$. With $E_w$ as given in (2),
this predicts a supernova remnant with
%\begin{eqnarray}
$E_{SNR}\simeq\onehalf\beta E_w\simeq 2\times 10^{51}\mbox{erg},$
%\end{eqnarray}
which is very similar to energies of non-GRB supernovae remnants. 
We emphasize that ultimately, this connection is to be applied
the other way around: obtaining estimates for $E_w$ from kinetic energies
in a sample of supernova remnants around black hole binaries, by
assuming that $\beta\sim0.1$ holds as a representative value for the
initial ejection velocity obtained from $E_w$. This assumption may be
eliminated by averaging over observed values of $\beta$ in a sample
of GRB-supernova events with identified line-doppler shifts.

\section{Remnants of beauty}

The angular momentum in rapidly spinning black holes produced in
core-collapse probably derives from orbital angular momentum,
following prior interaction of the young massive star with a binary 
companion (Paczynski 1998; Brown 2000). Therefore, the end product of 
the black hole-supernova is a black hole binary (a black hole and
a low mass companion) surrounded by a 
supernova remnant, accompanied by a burst of gravitational radiation 
as an echo in eternity.

It becomes of interest, therefore, to search for black hole binaries
in supernova remnants. A particularly striking example of an X-ray
binary surrounded by a supernova remnant is RX J050736-6847.8. 
The above suggests that this X-ray binary may harbor a black hole.

\section{Conclusions}

We present a mechanism for long-duration bursts (of tens of seconds) of
gravitational radiation powered by Kerr black holes. This emission
is catalyzed by a surrounding torus, in association with minor
outputs in winds, thermal and MeV neutrino emissions. A small
fraction of the black hole-spin energy is released contemporaneously
along the axis of rotation. We propose this as a model for GRB-supernovae
(``hypernovae"), following Woosley's (1993) scenario of black hole
formation in core-collapse of massive stars (Paczynski 1998, Brown 2000).

\begin{figure}
%\plotfiddle{f11.eps}{1.60in}{0}{40}{40}{-120}{00}
\plotone{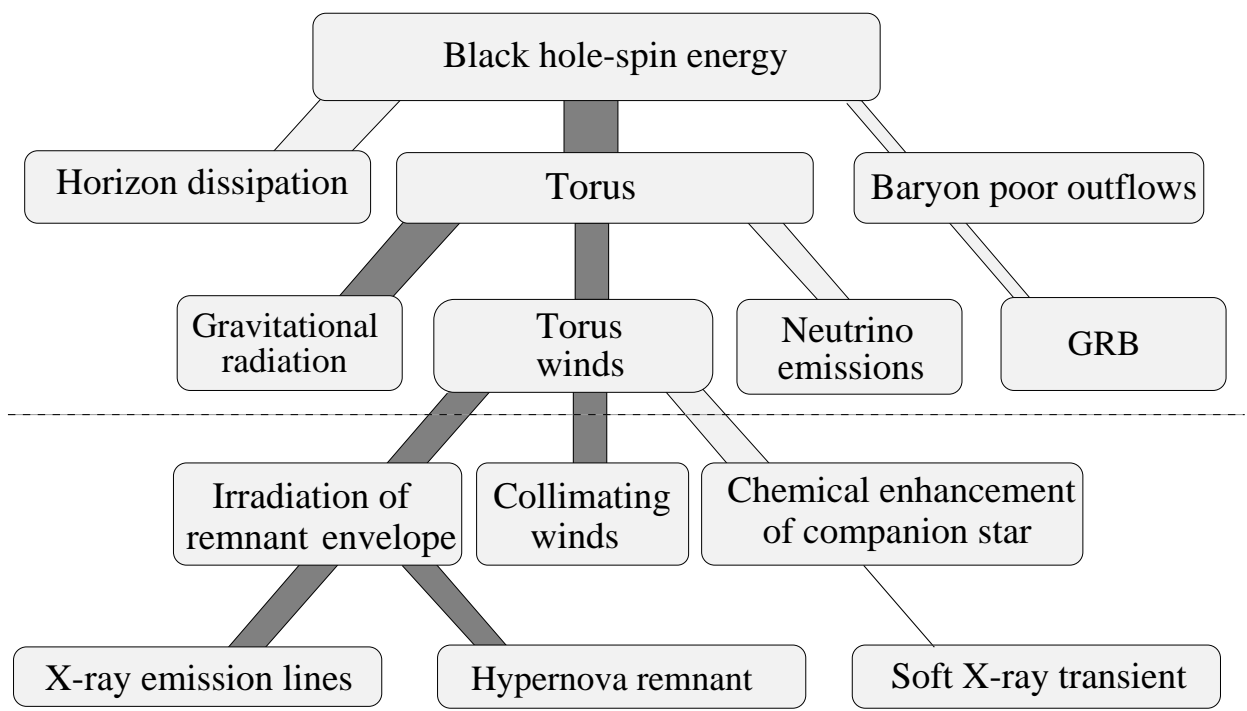}
\caption{A tree of black hole-spin energy, 
catalyzed by a torus: most energy is dissipated in the horizon, 
most of the output is in gravitational radiation, accompanied by a minor output 
in winds, thermal and neutrino emissions. A small fraction is released in 
baryon poor outflows as input to GRB-afterglow emissions. Direct measurement of 
gravitational radiation by upcoming gravitational wave-experiments provides a 
calorimetric test for Kerr black holes (dark connections).
Calorimetry on torus winds (below the dashed line, incomplete or 
unknown) provides an avenue for constraining the angular velocity 
of the torus, and hence its frequency of gravitational radiation.
Ultimately, this leaves a black hole binary, possibly an SXT, 
surrounded by an SNR. (Reprinted from van Putten \& 
Levinson, ApJ, 2003, in press.)}
\end{figure}

The proposed model accounts quantitatively for the secular timescale of 
long GRBs and the relatively small GRB-energies. The torus emissions in
various energy channels are correlated with the Keplerian angular velocity, 
which establishes a relation between wind-energies and the frequency of 
quadrupole gravitational radiation. This provides an avenue to use 
calorimetry on torus winds and its remnants in support of predicting
frequencies of gravitational radiation. This is intended for 
designing optimal detection
and search strategies in upcoming gravitational wave-experiments.
The interaction of the torus wind with the remnant evelope may further
produce a burst in radio emission, which could be a source of interest
for LOFAR/SKA. Some constraints might be placed on the statistics of 
these bursts by the recent analysis of Levinson et al. (2002).

The model predicts hypernova remnants in the form of black hole binaries
surrounded by a supernova remnant. A candidate of particular interest is
RX J050736-6847.8. It becomes of interest to pursue searches for a sample
of such X-ray binaries surrounded by supernova remnants, of which some may
harbor black hole binaries.

\acknowledgments
The author thanks G. Mendell for useful comments, and S. Kim for drawing attention
to RX J050736-6847.8. This research is supported by NASA Grant 5-7012, 
a NATO Collaborative Linkage Grant and an MIT C.E. Reed Fund. 
The LIGO Observatories were constructed by the California Institue of Technology
and Massachusetts Institute of Technology with funding from the National Science
Foundation under cooperative agreement PHY 9210038. The LIGO Laboratory operates
under cooperative agreement PHY-0107417. This paper has been assigned LIGO
Document Number LIGO-P030003-00-R.

\end{document}